\documentstyle[aps,epsf,rotate,citesort,preprint]{revtex}

\begin{document}

\draft

\title{The Web of Human Sexual Contacts}
\maketitle
           


Many ``real-world'' networks are clearly defined \cite{Strogatz01} while most
``social'' networks are to some extent subjective
\cite{Kochen89,Wasserman94}.  Indeed, the accuracy of empirically-determined
social networks is a question of some concern because individuals may have
distinct perceptions of what constitutes a social link.
One unambiguous type of connection is sexual contact.  Here we analyze data
on the sexual behavior of a random sample of individuals \cite{Lewin98}, and
find that the cumulative distributions of the number of sexual partners
during the twelve months prior to the survey decays as a power law with
similar exponents $\alpha \approx 2.4$ for females and males.
The scale-free nature of the web of human sexual contacts suggests that
strategic interventions aimed at preventing the spread of
sexually-transmitted diseases may be the most efficient approach.

%
%
%


Recent studies of real-world networks \cite{Strogatz01} have formalized
mathematically the ``six-degrees of separation'' concept put forth in the
classic study of Milgram \cite{Milgram67}.  This so-called small-world
phenomenon \cite{Watts98} refers to the surprising fact that networks have
small average path lengths between nodes while preserving a large degree of
``clustering'' \cite{Wasserman94}.  Small-world networks may belong to three
classes---single-scale, broad-scale, or scale-free---depending on their
connectivity distribution $P(k)$, where $k$ is the number of links connecting
to a node \cite{Amaral00}.  Scale-free networks---which are characterized by
a power law decay of the cumulative distribution, $P(k) \sim
k^{-\alpha}\,\,$--- may be formed due to preferential attachment, i.e., new
links are established preferentially between nodes with high connectivities
\cite{Simon55,Barabasi99}.



We analyze data gathered in a 1996 Swedish survey of sexual behavior
\cite{Lewin98}.  The survey---involving a random sample of 4781 Swedish
individuals (ages 18--74 yr)---used structured personal interviews and
questionnaires to collect information.  The response rate was 59 percent,
corresponding to 2810 respondents.  Two independent analyses of non-response
error reveal that elderly people, and especially elderly women, are
under-represented in the sample; apart from this skewness, the sample is
representative in all demographic dimensions.

\begin{figure}
\narrowtext
\centerline{
\epsfysize=0.70\columnwidth{\epsfbox{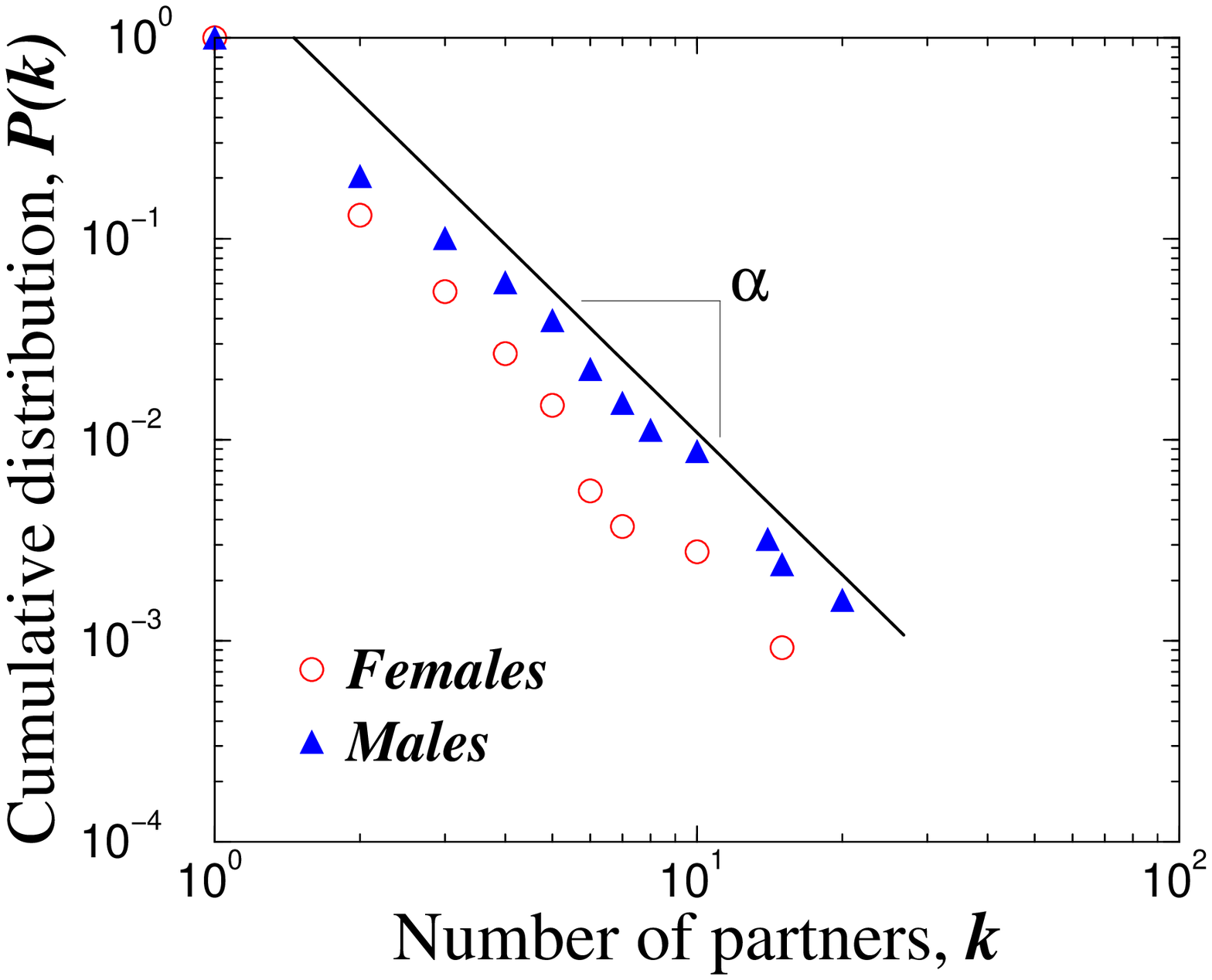}}}
\centerline{
\epsfysize=0.70\columnwidth{\epsfbox{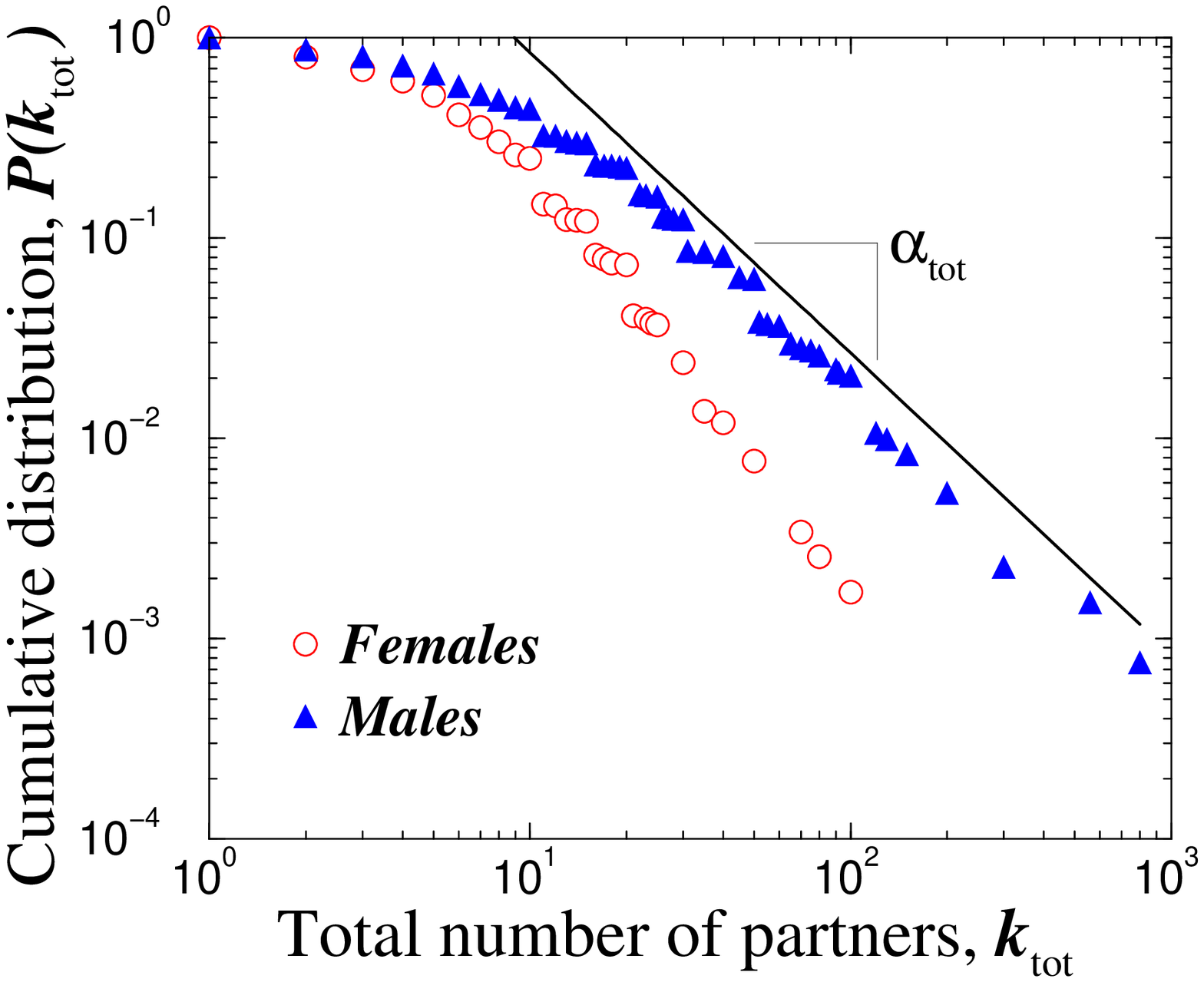}}
}
\vspace*{0.5cm}
\caption{ Scale-free distribution of number of sexual partners for females
and males.
{\bf a}, Distribution of number of partners $k$ in only the previous year.
Note the larger average number of partners for male respondents.  This
difference may be due to ``measurement bias''---social expectations may lead
males to ``inflate'' the number of sexual partners.  Note that the two
distributions are both linear, indicating power law scale-free behavior.
Moreover, the two curves are roughly parallel, indicating similar scaling
exponents.  For females, we obtain $\alpha = 2.54 \pm 0.2$ in the range
$k>4$, and for males, we obtain $\alpha = 2.31 \pm 0.2$ in the range $k>5$.
{\bf b}, Distribution of the total number of partners $k_{\mbox{\scriptsize
tot}}$ over all years since sexual initiation.  For females, we obtain
$\alpha_{\mbox{\scriptsize tot}} = 2.1 \pm 0.3$ in the range
$k_{\mbox{\scriptsize tot}} > 20$, and for males, we obtain
$\alpha_{\mbox{\scriptsize tot}} = 1.6 \pm 0.3$ in the range $20 <
k_{\mbox{\scriptsize tot}} < 400$. These two estimates agree within
statistical uncertainty.  }
\label{f-partners}
\end{figure}

Connections in the network of sexual contacts appear and disappear as sexual
relations are initiated and terminated.  To analyze the connectivity of this
dynamic network, whose links may be quite short lived, we first analyze the
number $k$ of sex partners over a relatively short time window---the twelve
months prior to the survey.  Figure~\ref{f-partners}a shows the cumulative
distribution $P(k)$ for both female and male respondents.  The data follow
closely a straight line in a double-logarithmic plot, consistent with a power
law dependence.  The data shows that males report a larger number of sexual
partners than do females \cite{Laumann94}, but that both have the same
scaling properties.

These results contrast with the exponential or Gaussian distributions---for
which there is a well-defined scale---as was recently found for friendship
networks \cite{Amaral00}.  Plausible mechanisms that could account for the
observed structure include: (i) increased skill in getting new partners as
the number of previous partners grows, (ii) different levels of
attractiveness, (iii) the need to have many new partners to maintain
self-image. Thus, the data are consistent with the preferential attachment
mechanism.  Perhaps, in sexual contact networks, as in other scale-free
networks, ``the rich do get richer \cite{Simon55,Barabasi99}.

We next analyze the total number $k_{\mbox{\scriptsize tot}}$ of partners in
the respondent's life up until the time of the survey.  This quantity is not
relevant to the ``instantaneous'' structure of the network but may help
elucidate the mechanisms responsible for the distribution of number of
partners.  Figure~\ref{f-partners}b shows the cumulative distribution
$P(k_{\mbox{\scriptsize tot}})$.  For values of $k_{\mbox{\scriptsize tot}} >
20$, the data follow a straight line in a double-logarithmic plot, consistent
with a power law dependence in the tails of the distribution.


Our major finding is the {\it scale-free\/} nature of the connectivity of an
objectively defined, non-professional, social network.  This result shows
that the concept of the ``core group'' considered in epidemiological studies
\cite{Hethcote84} is somewhat arbitrary as there is no well-defined threshold
or boundary separating the core group from other individuals (as there would
for a bimodal distribution).

Our findings also have possible epidemiological implications.  First,
epidemics arise and propagate much faster in scale-free networks than in
single-scale networks \cite{Watts98,Vespignani01}.  Second, measures to
contain or stop the propagation of diseases in a network must be radically
different for scale-free networks.  Specifically, the study of scale-free
networks indicates that they are resilient to random failure, but are highly
susceptible to destruction of the best connected nodes \cite{Albert00}, while
single-scale networks are not susceptible to attack even of the best
connected nodes.  Hence, the possibility that the web of sexual contacts has
a scale-free structure indicates that strategic targeting of safe-sex
education campaigns to those individuals with a large number of partners may
have a significant effect in reducing the propagation of sexually-transmitted
diseases.

We thank M. Buchanan, G. Franzese, S. Mossa, and C. M. Roman for helpful
suggestions and discussions, and especially G. Helmius for making the survey
data availble to us. FL and Y\AA ~thank STINT(97/1837) and HSFR(F0688/97),
and CE thanks HSFR(F0624/1999) for support.  LANA and HES thank NIH/NCRR (P41
RR13622) for support.



\noindent
{\bf Fredrik Liljeros$^{1}$, Christofer R. Edling$^1$, Lu\'{\i}s A.  Nunes
Amaral$^{2}$, H. Eugene Stanley$^2$, and Yvonne \AA berg$^1$} \\
\noindent
$^1$Department of Sociology, Stockholm University,
	   S-106 91 Stockholm, Sweden \\
\noindent
$^2$Center for Polymer Studies \& Dept. of Physics,
	   Boston University, Boston, MA 02215, USA


\end{document}